# Low $^{60}Fe$ abundance in Semarkona and Sahara 99555

Haolan Tang,[1,2*] Nicolas Dauphas[1]

[1]Origins Lab, Department of the Geophysical Sciences and Enrico Fermi Institute, The University of Chicago, 5734 South Ellis Avenue, Chicago IL 60637

[2]Ion Probe Group, Department of Earth and Space Sciences, University of California, Los Angeles, 595 Charles E. Young Drive East, Los Angeles CA, 90095

[*]To whom correspondence should be addressed. E-mail: haolantang@ucla.edu

**Abstract**

**Iron-60 ($t_{1/2}$=2.62 Myr) is a short-lived nuclide that can help constrain the astrophysical context of solar system formation and date early solar system events. A high abundance of $^{60}Fe$ ($^{60}Fe/^{56}Fe \approx 4\times10^{-7}$) was reported by *in situ* techniques in some chondrules from the LL3.00 Semarkona meteorite, which was taken as evidence that a supernova exploded in the vicinity of the birthplace of the Sun. However, our previous MC-ICPMS measurements of a wide range of meteoritic materials, including chondrules, showed that $^{60}Fe$ was present in the early solar system at a much lower level ($^{60}Fe/^{56}Fe \approx 10^{-8}$). The reason for the discrepancy is unknown but only two Semarkona chondrules were measured by MC-ICPMS and these had Fe/Ni ratios below ~2× chondritic. Here, we show that the initial $^{60}Fe/^{56}Fe$ ratio in Semarkona chondrules with Fe/Ni ratios up to ~24× chondritic is $(5.39\pm3.27)\times10^{-9}$. We also establish the initial $^{60}Fe/^{56}Fe$ ratio at the time of crystallization of the Sahara 99555 angrite, a chronological anchor, to be $(1.97\pm0.77)\times10^{-9}$. These results demonstrate that the initial abundance of $^{60}Fe$ at solar system birth was low, corresponding to an initial $^{60}Fe/^{56}Fe$ ratio of $(1.01\pm0.27)\times10^{-8}$.**

**Keywords: Planetary Systems: meteorites —— methods: analytical —— Planetary Systems: protoplanetary disks —— ISM: abundances**

## 1. Introduction

Chondrules are quenched spherical beads that were once molten in space and are found in large abundance in primitive meteorites known as chondrites (Scott and Krot, 2014). Although the mechanism responsible for their melting is uncertain (shock waves, planetary collisions, or lightning), they have been precisely dated using several radioactive chronometers. Their crystallization ages span a few million years with a peak at ~2 Myr after formation of calcium-aluminum-rich inclusions (CAIs) (Hutcheon and Hutchison, 1989; Kita et al., 2000; Rudraswami et al., 2008; Villeneuve et al., 2009; Kita and Ushikubo, 2012), which is taken to represent time zero in early solar system chronology (Dauphas and Chaussidon, 2011). Chondrules formed early and contain some phases that have high Fe/Ni ratios, which makes them particularly well suited to establish the abundance of $^{60}$Fe in the early solar system.

Using multi-collector inductively coupled plasma mass spectrometry (MC-ICPMS), Tang and Dauphas (2012) found a low and uniform initial $^{60}$Fe/$^{56}$Fe ratio in early solar system materials, including in chondrites and their constituents. However, other *in situ* measurements by secondary ionization mass spectrometry (SIMS) have yielded much higher $^{60}$Fe/$^{56}$Fe ratios (Telus et al., 2013 and references therein; Mishra and Goswami, 2014; Mishra and Chaussidon, 2014). Several explanations to this discrepancy are possible: (1) SIMS measurements suffer from an unidentified isobaric interference on $^{60}$Ni, (2) $^{60}$Fe was heterogeneously distributed and the $^{60}$Fe/$^{56}$Fe ratio is highly variable from chondrule-to-chondrule, or (3) parent-body alteration and metamorphism have disturbed $^{60}$Fe-$^{60}$Ni systematics in chondrules. Semarkona is the material of choice to study $^{60}$Fe-decay because this LL3.00 ordinary chondrite has experienced little thermal metamorphism (Quirico et al., 2003; Grossman and Brearley, 2005) with a peak temperature probably not exceeding ~200 °C (Huss et al., 2006), so disturbance to the $^{60}$Fe-$^{60}$Ni system should be minimal. A limitation to studying this sample is its availability, as its total mass is only 691 g and less than of third of this is available in meteorite collections for scientific studies. In Tang and Dauphas (2012), we analyzed 2 bulk chondrules from Semarkona, 8 chondrules and magnetic/size separates from NWA 5717 (an ungrouped ordinary chondrite with a petrologic type of 3.05, *i.e.*, slightly more metamorphosed than Semarkona). The spread in Fe/Ni ratio was limited (up to ~5× chondritic) but this was sufficient to set an upper-limit on the initial $^{60}$Fe/$^{56}$Fe ratio of $<3\times10^{-8}$, clearly lower than the values found by SIMS of $\sim4\times10^{-7}$. More recently, Telus et al. (2013) measured Fe and Ni distribution in chondrules by synchrotron X-ray fluorescence and found that late-stage fluids had mobilized these elements in most chondrites. However, only 5/16 Semarkona chondrules showed evidence of mobilization, meaning that most chondrules in that meteorite (~69 %) should be relatively pristine. The chondrules affected by parent-body disturbance should have relatively low Fe/Ni ratios.

In an effort to better constrain the abundance of $^{60}$Fe in the early solar system, we have analyzed 6 new chondrules from Semarkona, some of which have high Fe/Ni ratios (up to ~24× chondritic). We have also studied the initial $^{60}$Fe/$^{56}$Fe ratio at the time of crystallization of the quenched angrite Sahara 99555, a sample that has been dated by several techniques (Connelly et al., 2008b; Spivak-Birndorf et al., 2009; Schiller et al., 2010; Larsen et al., 2011; Kleine et al., 2012), and can serve as a chronological anchor to back-calculate the initial $^{60}$Fe/$^{56}$Fe ratio at the time of CAI formation. We confirm our earlier conclusion that the initial $^{60}Fe/^{56}Fe$ ratio was uniform across the inner protoplanetary disk at a level of $10^{-8}$.

In Sect. 2, the samples and their processing are presented. This includes retrieval of chondrules from the Semarkona meteorite; magnetic, density, and size separations of grains from the Sahara 99555 angrite; characterization of the samples, digestion, purification of Ni by chromatography, and isotopic analysis by mass spectrometry. In Sect. 3, the Ni isotopic results are presented, and the implications of those measurements for the abundance of $^{60}Fe$ in the chondrule-forming region and quenched angrites are discussed in Sect. 4. Sect.5 concludes that the abundance of $^{60}Fe$ was low; implying that $^{26}Al$ in meteorites came from the winds of one or several massive stars while $^{60}Fe$ was inherited from galactic background.

## 2. Samples and Methods

### 2.1. Semarkona chondrules and Sahara 99555 angrite

Chondrules are quenched droplets of magma that formed within a few million years of the formation of the solar system (Kita and Ushikubo, 2012; Scott and Krot, 2014). After incorporation in chondrite parent-bodies, they were subjected to thermal metamorphism and aqueous alteration, processes that could disturb $^{60}Fe$-$^{60}Ni$ systematics, so care must be exercised in selecting the most pristine samples for study. Semarkona is a LL3.00 chondrite (Quirico et al., 2003; Grossman and Brearley, 2005), meaning that it was minimally modified by parent-body aqueous alteration or metamorphism (the degree of aqueous alteration increases from type 3 to 1 while the degree of metamorphism increases from type 3 to 6). The Smithsonian Institution provided a fragment of Semarkona of approximately ~ 370 mg, from which 14 chondrules were hand-picked under a binocular microscope. In order to identify the chondrules with relatively high Fe/Ni ratios, small areas were polished using 1200 grit abrasive paper. The internal areas thus exposed allowed us to measure the chemical compositions of the chondrules using energy dispersive spectroscopy on a JEOL JSM-5800LV scanning electron microscope operated in low vacuum mode to prevent charging. The surfaces were not well polished and the samples were not carbon coated, so the chemical analyses were of limited quality but sufficient for our purposes. Out of the 14 starting chondrules, 6 had relatively high Fe contents and were selected for Ni isotopic analysis. The bulk chondrules (~2-14 mg) were first rinsed with acetone to get rid of possible surface contamination. To avoid the risk of accidental sample loss, recover the maximum amount of Ni for high-precision isotopic analysis (a low $^{60}Fe/^{56}Fe$ ratio implies that $^{60}Ni$-excess should be barely resolvable in chondrules), and preserve the bulk nature of the measurements, no further characterization or fragmentation was done on these samples, which were directly digested.

Angrites are a group of basaltic achondrites that record some early igneous activity (Mittlefehldt et al., 2002; Mittlefehldt, 2003; Keil, 2012). According to their textural characteristics, angrites are divided into two subgroups. Fine-grained angrites, such as D'Orbigny and Sahara 99555, are characterized by quenched textures indicative of rapid cooling (Mittlefehldt et al., 2002). Coarser-grained angrites, such as Angra dos Reis and NWA 4801, experienced more protracted cooling histories (*e.g.*, Nyquist et al., 2009; Kleine et al., 2012). High-resolution chronometers, including $^{26}Al$-$^{26}Mg$, $^{53}Mn$-$^{53}Cr$, $^{182}Hf$-$^{182}W$, and Pb-Pb, have been applied to date angrites (Lugmair and Galer, 1992; Lugmair and Shukolyukov, 1998; Nyquist et al., 1994, 2003; Baker et al., 2005; Amelin, 2007, 2008; Markowski et al., 2007; Connelly et al., 2008a,b; Spivak-Birndorf et al., 2009), providing a means of testing the concordance between different extant and extinct radiochronometers. Due to its quenched texture and rapid cooling, Sahara 99555 is a well-suited anchor for early solar system chronology. It is mainly composed of Ca-rich olivine (~31-42 %), Al-Ti rich pyroxene (~24-28 %), and anorthitic plagioclase (~33-39 %) (Mikouchi et al., 2000; Mikouchi and McKay 2001). A 500 mg sample of Sahara 99555 purchased from L. Labenne was crushed into powder in an agate mortar. The fragmented samples were separated into two parts: one was processed with a hand magnet followed by sieving to separate the grains into three silicate size fractions (100-166 μm, 166-200 μm and >200 μm); the other was processed with a hand magnet and split according to density (below or above 3.10 g/cm$^3$) using sodium polytungstate solution. The mineral fractions were not characterized but Tang and Dauphas (2012) applied the same procedure to D'Orbigny angrite and the low-density fraction was relatively rich in anorthite while the high-density fraction was rich in olivine and pyroxene. For density separation, the fragmented sample was placed in the separatory funnel and heavy liquid was then added. The funnel was left for 10 minutes until dense grains (>3.10 g/cm$^3$) sank to the bottom while light grains (<3.10 g/cm$^3$) remained suspended in the liquid. Silicate fractions with different densities were collected onto pieces of weighing paper, rinsed with Millipore Milli-Q water, and dried in an oven.

To assess data quality and make sure that no analytical artifact was present, terrestrial standards were processed and analyzed together with the meteorite samples.

### 2.2. Sample preparation, digestion, and chemical separation

The chemistry was performed under clean laboratory conditions at the Origins Lab of the University of Chicago. Optima grade HF, reagent grade acetone, and double distilled HCl and $HNO_3$ were used for digestion and column chromatography. Millipore Milli-Q water was used for acid dilution.

Tang and Dauphas (2012, 2014) provide details on the procedures of sample digestion and chemical separation. Semarkona chondrules and Sahara 99555 fractions weighing 2 to 130 mg were digested in 5-20 ml HF-$HNO_3$ (in a 2:1 volume ratio) in a Teflon beaker placed on a hot plate at ~90 ˚C for 2 days. The solution was subsequently evaporated to dryness and re-dissolved in a 5-20 mL mixture of concentrated HCl-$HNO_3$ (in a 2:1 volume ratio) until all the sample powder was completely digested. The solutions were dried down and the residues taken back in solution with a minimum amount of concentrated HCl (~11 M) for loading on the first column. In order to obtain sufficiently clean Ni for isotopic measurements, chemical separation of Ni from matrix elements and isobars was done in three steps of chromatography that are described below.

U/TEVA cartridge (Horwitz et al., 1992) was used for the first step of chemistry to get rid of Ti, Co, Zr and Fe. The column (2 mL volume, 2.5 cm length, 1 cm diameter) was pre-cleaned with 10 mL water, 15 mL 0.4 M HCl, 15 mL 4 M HCl and was then conditioned with 10 mL of concentrated HCl. The sample solution was loaded onto the column in 5-10 mL 10 M HCl. The load solution was collected in clean Teflon beakers and an additional 10 mL of concentrated HCl was passed through the resin and collected in the same beaker. This eluate contained Ni together with Na, Mg, Ca and other matrix elements. After drying down, the Ni elution cut from the first column chemistry was re-dissolved in 5 mL of a mixture of 20 % 10 M HCl-80 % acetone (by volume) and loaded onto a Teflon column containing 5 mL (40 cm length, 0.4 cm diameter) of pre-cleaned Bio-Rad AG50-X12 200-400 mesh hydrogen-form resin, previously conditioned with 10 mL 20 % 10 M HCl-80 % acetone. After loading the sample solution and rinsing with 30 mL 20 % 10 M HCl – 80% acetone mixture to eliminate Cr and any remaining Fe, Ni was collected by eluting 150 mL of the HCl-acetone mixture into a jar containing 30 mL of water to dilute HCl and stabilize Ni in the eluate. In those conditions, Mg, Na, Ca, and other matrix elements were retained on the resin (Strelow et al., 1971). The collected Ni solution was evaporated at moderate temperature (<90 °C) under a flow of $N_2$ to avoid the formation of organic complexes with acetone and accelerate evaporation. After evaporation, the Ni fraction was dissolved in 1 mL of aqua regia (1:3 $HNO_3$:HCl) to remove any organic residue formed during evaporation. This HCl-acetone column was repeated five times to ensure thorough separation of major rock forming element Mg from Ni, two elements that are notoriously difficult to separate. Zinc is a significant interference on low abundance isotope $^{64}$Ni. It was removed using a third column filled with 1 mL (2 cm length, 0.8 cm diameter) Bio-Rad AG1W-X8 anionic ion exchange resin in 8 M HBr medium (Moynier et al., 2006). Nickel was eluted in 8 mL 8 M HBr, whereas Zn was retained on the resin.

The entire procedural blank was ~35 ng, which is negligible compared to the amounts of Ni in the samples. The nickel yield of the entire procedure was 90-100 %.

### 2.2 Mass spectrometry

All measurements were performed at the Origins Laboratory of the University of Chicago using a Neptune MC-ICPMS equipped with an OnTool Booster 150 (Pfeiffer) interface jet pump. Jet sampler and X skimmer cones were used. The samples were re-dissolved in 0.3 M $HNO_3$ and introduced into the mass spectrometer with Ar + $N_2$ using an Aridus II desolvating nebulizer at an uptake rate of ~ 100 μL/min. The instrument sensitivity for $^{58}$Ni was 100 V/ppm. One analysis consisted of 25 cycles, each acquisition lasting for 8.4 s. During a session, each sample solution was measured 13 times bracketed by SRM 986. A small isobaric interference from the least abundant isotope of iron, $^{58}$Fe, on the most abundant isotope of nickel, $^{58}$Ni, was corrected by monitoring $^{57}$Fe (the correction is always smaller than 0.5 ε). All isotopes were measured using Faraday cups with $10^{11}$ Ω resistance amplifiers. Background was subtracted using an on-peak zero procedure. Internal normalization was used to correct mass-dependent isotopic fractionation by fixing $^{61}$Ni/$^{58}$Ni to 0.016720 or $^{62}$Ni/$^{58}$Ni to 0.053389 (Gramlich et al., 1989) using the exponential law (Maréchal et al., 1999). Nickel-64 is reported only for samples that gave ~15 μg Ni because below this level, an isobaric interference from $^{48}Ti^{16}O^{+}$ can affect the results (Tang and Dauphas, 2012).

Approximately 20 % of the original sample solutions were kept as safety aliquots and for Fe/Ni ratio measurements by MC-ICPMS using both standard bracketing and standard addition techniques. The Fe/Ni ratios measured by standard addition agree well with Fe/Ni ratios measured by simple sample-standard bracketing. Fe/Ni ratios in terrestrial standards were all within 3 % of their reference values, demonstrating the accuracy of our measurements.

## 3. Results

Table 1 shows the Ni isotopic compositions and Fe/Ni ratios measured in Semarkona chondrules, Sahara 99555 fractions and terrestrial standards. Terrestrial standards passed through the same column chemistry as meteoritic samples have normal Ni isotopic ratios, attesting to the accuracy of the measurements. The intercepts and slopes of the $\varepsilon^{60}$Ni *vs.* $^{56}$Fe/$^{58}$Ni correlations were calculated using Isoplot (Ludwig, 2012) to estimate the initial $^{60}$Fe/$^{56}$Fe ratios and $\varepsilon^{60}$Ni values based on the following isochron equation (see Fig. 2 of Dauphas and Chaussidon, 2011 for an explanation),

$$\varepsilon^{60}\mathrm{Ni}_{\mathrm{present}} = \varepsilon^{60}\mathrm{Ni}_0 + 2.596\times10^4 \left(\frac{^{60}\mathrm{Fe}}{^{56}\mathrm{Fe}}\right)_0 \left(\frac{^{56}\mathrm{Fe}}{^{58}\mathrm{Ni}}\right)_{\mathrm{present}}, \quad (1)$$

where 2.596 is a constant that corresponds to the $^{58}$Ni/$^{60}$Ni ratio in solar system material. No significant variations were detected for $^{61}$Ni, $^{62}$Ni and $^{64}$Ni isotopes relative to terrestrial standards.

A total of eight Semarkona chondrules (6 out of 14 chondrules surveyed from this study, Table 1; 2 from Tang and Dauphas, 2012) have been measured and the results are shown in Fig. 1A. Fe/Ni ratios range from 13 to 435 (for reference, the CI chondrite value is 17). No significant $^{60}$Ni excess was detected in the chondrules with relatively low Fe/Ni ratios (Fe/Ni<100). One Type II chondrule (FeO ~ 16.7 wt%), SC-13-6, has a high Fe/Ni ratio (~435) as well as barely resolvable $\varepsilon^{60}$Ni excess of +0.051±0.043. Combining data from all Semarkona chondrules, a single isochron can be defined corresponding to initial $^{60}$Fe/$^{56}$Fe = $(5.39\pm3.27)\times10^{-9}$ and $\varepsilon^{60}$Ni = -0.032±0.023 (MSWD = 0.26), at the time of equilibration (Fig. 1A). This initial $^{60}$Fe/$^{56}$Fe ratio is much lower than the value inferred by SIMS of $\sim4\times10^{-7}$ at the time of chondrule formation (Mishra and Goswami, 2014; Mishra and Chaussidon, 2014). The value of the slope is heavily leveraged by SC-13-6. Even if this data point is excluded from the regression, the initial $^{60}$Fe/$^{56}$Fe ratio calculated based on MC-ICPMS data remains low, $^{60}$Fe/$^{56}$Fe = $(1.77\pm2.77)\times10^{-8}$, and is inconsistent with SIMS results.

The Fe-Ni results for mineral separates in the quenched angrite Sahara 99555 are given in Table 1. The mineral separates display high Fe/Ni ratios ranging from ~ 2,500 (for the low-density fraction) to ~6,900 (for the high-density fraction) and radiogenic $\varepsilon^{60}$Ni values between +0.21 and +0.54. The initial $^{60}Fe/^{56}Fe$ ratio and $\varepsilon^{60}$Ni value inferred from mineral separates from the Sahara 99555 angrite are shown in Fig. 1B. The data points define an internal isochron (MSWD = 0.49) of slope $^{60}Fe/^{56}Fe = (1.96\pm0.77)\times10^{-9}$ and intercept $\varepsilon^{60}Ni = +0.043\pm0.115$ at the time of closure to isotope exchange of the minerals investigated. This value agrees well with independent results reported for this meteorite, which gave an initial $^{60}Fe/^{56}Fe$ ratio of $(1.8\pm0.5)\times10^{-9}$ (Quitté et al., 2010).

## 4. Discussion

### 4.1. Abundance of $^{60}$Fe in the chondrule-forming region

The nickel isotopic compositions of Semarkona chondrules (LL3.00) were analyzed and give an initial $^{60}Fe/^{56}Fe$ ratio of $(5.39\pm3.27)\times10^{-9}$ at the time of chondrule formation. Telus et al. (2013, 2014) measured Fe and Ni distribution in chondrules by synchrotron X-ray fluorescence and found that late-stage fluids had mobilized these elements from the matrix to deposit them as iron and nickel oxide/hydroxide in chondrule fractures. Such mobilization has little bearing on the inferred low $^{60}$Fe abundance in Semarkona chondrules measured in this study for the following three reasons:

(*i*) In chondrites of metamorphic grade as low as 3.10, Telus et al. (2014) found that 100 % of the chondrules were affected by Fe-Ni mobilization. However, in Semarkona (type LL3.00, one of the most pristine meteorite samples available), they report that approximately 70 % of the chondrules analyzed did not display any evidence for mobilization (the probability to have a random Semarkona chondrule free of Fe-Ni mobilization is $p = 0.7$). The probability that $k$ out of $n = 8$ random Semarkona chondrules be free of Fe -Ni mobilization is,

$$P(k) = \binom{n}{k} p^k (1-p)^{n-k}. \qquad (2)$$

The calculated probabilities are $P(0)\sim 0\,\%$, $P(1)\sim 0\,\%$, $P(2)\sim 1\,\%$, $P(3)\sim 5\,\%$, $P(4)\sim 13\,\%$, $P(5)\sim 25\,\%$, $P(6)\sim 30\,\%$, $P(7)\sim 20\,\%$, and $P(8)\sim 6\,\%$. At 95 % confidence level, the majority ($k > 4$) of the chondrules analyzed here were not affected by Fe-Ni mobilization.

(*ii*) In bulk measurements, the effect of Fe-Ni mobilization would be to lower the Fe/Ni ratios and $\varepsilon^{60}$Ni values towards chondritic values. Such physical admixture of Fe-Ni in chondrules through fractures could have added some scatter to the data points and could have brought the samples towards the chondritic value but overall, the points should have moved along the isochron line. Indeed, mixing between two components in the $\varepsilon^{60}$Ni *vs*. Fe/Ni diagram is a straight line, so the isochronous behavior of bulk chondrule measurements should have been preserved at some level.

(*iii*) The Semarkona chondrules most likely to have been affected by Fe-Ni addition should be among those that display low Fe/Ni ratios in bulk. Chondrules with high Fe/Ni ratios such as SC-13-6 (Fe/Ni ratio ~24× chondritic) provide the most leverage to define the slope of the isochron and are least likely to have been affected by Fe/Ni mobilization. For reference, an initial $^{60}$Fe/$^{56}$Fe ratio of $4\times10^{-7}$ in Semarkona chondrules (Mishra and Goswami, 2014; Mishra and Chaussidon, 2014) should have been associated with excess $\varepsilon^{60}$Ni of +6.3 in chondrule SC-13-6 (Eq. 1) while a value of +0.051±0.043 was measured.

The results presented here thus reaffirm our earlier conclusion (Tang and Dauphas, 2012) that $^{60}$Fe was present at a low level in the chondrule-forming region, $^{60}$Fe/$^{56}$Fe=(5.39±3.27)×10$^{-9}$.

**4.2. Abundance of $^{60}$Fe in the Sahara 99555 angrite**

The mineral separates in the Sahara 99555 angrite give an initial $^{60}$Fe/$^{56}$Fe ratio of (1.96±0.77) ×10$^{-9}$ (Fig. 1B), which is in excellent agreement with the initial value of (1.8±0.5)×10$^{-9}$ reported by Quitté et al. (2010) in the same meteorite. The weighted average of those two values is (1.85±0.42)×10$^{-9}$.

The Ni isotopic compositions of mineral separates from the D'Orbigny meteorite, a quenched angrite like Sahara 99555, were measured independently by Quitté et al. (2010), Spivak-Birndorf et al. (2011) and Tang and Dauphas (2012). The initial $^{60}$Fe/$^{56}$Fe ratios reported by these three studies are (4.1±2.6)×10$^{-9}$, (2.81±0.86) ×10$^{-9}$ and (3.42±0.58)×10$^{-9}$, respectively. The three values agree and the weighted average is (3.26±0.47)×10$^{-9}$.

The calculated initial $^{60}$Fe/$^{56}$Fe ratio of D'Orbigny is significantly higher than the value measured in Sahara 99555. The age difference between these two angrites is given by $\Delta t_{2-1} = \ln(r_1/r_2)/\lambda$ with an associated error of $\sigma(\Delta t_{2-1}) = \sqrt{\sigma_{r_1}^2/r_1^2 + \sigma_{r_2}^2/r_2^2}/\lambda$, where $r$ denote the initial $^{60}$Fe/$^{56}$Fe ratio in either Sahara 99555 or D'Orbigny, and $\lambda$ is the half-life of $^{60}$Fe (2.62 Myr). The calculated age difference between D'Orbigny and Sahara 99555 is +2.1±1.0 Myr.

The relative chronology of formation of D'Orbigny and Sahara 99555 can be compared with independent estimates obtained using various dating techniques. Internal $^{26}$Al-$^{26}$Mg isochrons in Sahara 99555 and D'Orbigny give initial $^{26}$Al/$^{27}$Al ratios of (4.50±0.54)×10$^{-7}$ and (3.97±0.26)×10$^{-7}$, respectively (Spivak-Birndorf et al., 2009;

Schiller et al., 2010). These two values are almost indistinguishable, meaning that the two objects crystallized within ~0.2 Myr of each other. Similarly, $^{182}$Hf-$^{182}$W internal isochrons in Sahara 99555 and D'Orbigny give initial ratios of $(6.83\pm0.14)\times10^{-5}$ and $(7.15\pm0.17)\times10^{-5}$ (Kleine et al., 2012), corresponding to a time difference between D'Orbigny and Sahara 99555 of +0.6±0.4 Myr. Pb-Pb ages have also been reported for D'Orbigny (4563.37±0.25 Myr; Brennecka and Wadhwa, 2012) and Sahara 99555 (4563.64±0.14 Myr, Connelly et al., 2008b; Larsen et al., 2011). These absolute Pb-Pb ages should be regarded with caution because inter-laboratory calibration for this dating system is lacking, yet the ages are indistinguishable. All available evidence thus suggests that D'Orbigny and Sahara 99555 crystallized at approximately the same time.

The difference between $^{60}$Fe-$^{56}$Fe and other decay systems cannot be explained by a difference in closure age because the cooling rates of quenched angrites estimated from diffusion profiles and petrographic textures are rapid, 7-50 °C/hr (Mikouchi et al., 2001). This means that it would take 1 to 6 days for the samples to cool by 1000 °C, which covers more than the span of closure temperatures for the systems discussed above. This is instantaneous in regard to planetary timescales, so one would expect all extant and extinct chronometers to be closed to isotopic exchange at approximately the same time.

Most likely, the low $^{60}$Fe/$^{56}$Fe ratio measured in Sahara 99555 *vs*. D'Orbigny reflects terrestrial contamination in the former. Sahara 99555 was found in the Sahara desert. As pointed out by Crozaz et al. (2003), Floss et al. (2003), and Amelin (2008), Sahara 99555 shows more evidence of terrestrial alteration than other angrites, including mobilization of rare earth elements. The sample that we studied was partly covered with some desert-weathering product that we physically removed but Crozaz et al. (2003) and Floss et al.

(2003) showed that chemical alteration also affected Sahara 99555 samples with fresh appearances. The $^{60}Fe/^{56}Fe$ ratio of $(3.26\pm0.47)\times10^{-9}$ from D'Orbigny internal isochrons may thus provide the best estimate of the $^{60}Fe/^{56}Fe$ ratio at the time of crystallization of the quenched angrites (Quitté et al., 2010; Tang and Dauphas, 2012).

### 4.3. Abundance of $^{60}$Fe at solar system birth and astrophysical implications

Tang and Dauphas (2012) were able to constrain the $^{60}Fe/^{56}Fe$ ratio at solar system birth to $(1.15\pm0.26)\times10^{-8}$ using internal isochrons in Gujba CB chondrite, D'Orbigny angrite, and unequilibrated ordinary chondrites as well as bulk rock isochrons for angrites and HED meteorites. The new results presented here for Semarkona chondrules and Sahara 99555 allow us to refine the initial $^{60}Fe/^{56}Fe$ ratio. Table 2 gives the ages relative to CAIs and initial $^{60}Fe/^{56}Fe$ ratios for all MC-ICPMS measurements for which resolvable $^{60}$Ni-excess could be resolved, that is to say Semarkona chondrules, bulk HEDs, bulk angrites, and mineral separates of D'Orbigny and Sahara 99555. Those data relate to the $^{60}Fe/^{56}Fe$ ratio at the time of CAI formation using the free decay equation,

$$\left({}^{60}\mathrm{Fe}/{}^{56}\mathrm{Fe}\right)_{\mathrm{t}} = \left({}^{60}\mathrm{Fe}/{}^{56}\mathrm{Fe}\right)_{\mathrm{t_{CAI}}} e^{-\lambda(\mathrm{t}-\mathrm{t_{CAI}})}, \quad (3)$$

where $\mathrm{t}$ and $\mathrm{t_{CAI}}$ are the formation ages of the samples considered and CAIs, respectively. The $^{60}Fe/^{56}Fe$ ratios inferred from $\varepsilon^{60}Ni$-$^{56}Fe/^{58}Ni$ isochrons are plotted in Fig. 2 and all the results obtained thus far point to an initial $^{60}Fe/^{56}Fe$ ratio of $(1.01\pm0.27)\times10^{-8}$ and a homogeneous distribution of $^{60}$Fe (see Table 2).

The results from the present study reaffirm and strengthen our earlier conclusion that the abundance of $^{60}$Fe in the early solar system was low. This contrasts with recent *in situ* studies of chondrules from unequilibrated ordinary chondrites that report high initial $^{60}Fe/^{56}Fe$ ratios. Telus et al. (2014) made the case that most chondrites other than Semarkona have been affected by Fe-Ni mobilization. For this reason, we focus our comparison on the results obtained by SIMS on chondrules from the Semarkona meteorite. The initial $^{60}Fe/^{56}Fe$ ratios reported by Mishra and Goswami (2014) and Mishra and Chaussidon (2014) correspond to an initial $^{60}Fe/^{56}Fe$ ratio at solar system birth of ~$70\times10^{-8}$. This is almost two orders of magnitude higher than the initial $^{60}Fe/^{56}Fe$ ratio obtained by MC-ICPMS (~$10^{-8}$, Tang and Dauphas, 2012; this study). This cannot be due to $^{60}Fe$ heterogeneity because Tang and Dauphas (2012) did not detect isotopic anomalies in $^{58}Fe$, which is produced in stars together with $^{60}Fe$. Furthermore, the same sample types, Semarkona chondrules, were measured by both SIMS and MC-ICPMS. Telus et al. (2013) also reported SIMS measurements of 3 Semarkona chondrules and found barely detectable $^{60}Fe$ in only one chondrule, $^{60}Fe/^{56}Fe$ ~$(1.4\pm1.2)\times10^{-7}$, while the other two chondrules only provided upper-limits. One such chondrule has an upper-limit of $^{60}Fe/^{56}Fe$ $<5.1\times10^{-8}$, which is significantly lower than the values reported by Mishra and Goswami (2014) and Mishra and Chaussidon (2014) (Fig. 3).

We have no explanation for the discrepancy between the MC-ICPMS and SIMS studies but we note that all SIMS estimates are based on measurements of a single sample type (*i.e.*, pyroxene in chondrules from unequilibrated ordinary chondrites) and that most of the SIMS isochrons are defined by points that have error bars that largely overlap with zero, the significance of the isochron arising from a few data points with small errors (Fig. 3). In contrast, the estimate from MC-ICPMS is derived from isochrons measured on a variety of samples formed at different times (bulk HEDs –Vesta, SNCs –Mars, bulk angrites, D'Orbigny and Sahara 99555 angrites, Gujba, NWA 5717, and Semarkona chondrules) and the significances of the regressions are in most cases very well established (Fig. 1B; Tang and Dauphas, 2012). The isochron for Semarkona chondrules presented here (Fig. 1A) is barely resolvable from zero but the corresponding initial $^{60}Fe/^{56}Fe$ ratio is much lower than the initial ratio expected based on SIMS data. Care was put in the SIMS studies to avoid analytical artifacts and it is not clear how much improvement can be made on that front but other instruments exist or are being developed that may provide a direct comparison with SIMS measurements, such as resonant ionization mass spectrometry, atom probe, or megaSIMS. At present, the weight of evidence supports a low and uniform abundance of $^{60}Fe$ in the early solar system.

The low initial $^{60}Fe/^{56}Fe$ ratio is consistent with background abundances in the Galaxy with no compelling need to invoke late injection from a nearby star (Tang and Dauphas, 2012). Indeed, the average $^{60}Fe/^{56}Fe$ ratio in the ISM at solar system birth inferred from γ-ray astronomy (Wang et al, 2007) is $(2.8\pm1.4)\times10^{-7}$, which is 30 times higher than the initial ratio in the early solar system (Tang and Dauphas, 2012). For comparison, the average $^{26}Al/^{27}Al$ ratio in the ISM at solar system birth inferred from γ-ray astronomy (Diehl et al., 2006, 2010) is $(3.0\pm0.8)\times10^{-6}$, which is 17 times lower than the early solar system ratio (Lee et al., 1976). It thus appears that $^{26}Al$ and $^{60}Fe$ in meteorites have different origins (Tang and Dauphas, 2012).

Several scenarios can be considered to incorporate freshly made $^{26}Al$ without adding too much $^{60}Fe$. Adjusting the timescale between nucleosynthesis and solar system formation does not help because $^{26}Al$ has a shorter half-life than $^{60}Fe$, so any delay would cause the $^{26}Al/^{60}Fe$ ratio to decrease, making the problem worse. The possible scenarios to explain the high $^{26}Al/^{60}Fe$ ratio of the early solar system include (1) supernova explosion with fallback of the inner layers, so that only $^{26}Al$ is efficiently ejected while

$^{60}Fe$ is trapped in the stellar remnant (Meyer et al., 2000) but this is unlikely because one would need to have a lot of fallback (a cut-off in the C/O-burning layer) to prevent $^{60}Fe$ from escaping (Takigawa et al., 2008), (2) interaction of a supernova with an already formed cloud core, so that only the outer layers are efficiently injected while the inner layers are deflected (Gritschneder et al., 2012), and (3) ejection of $^{26}Al$ as winds from one or several massive stars (Arnould et al., 1997, 2006 ; Gaidos et al., 2009; Tatischeff et al., 2010; Gounelle et al., 2012; Young, 2014). The last scenario is appealing because it could be a natural outcome of the presence in the solar system forming region of one or several Wolf-Rayet (W-R) stars, as such stars shed their mass through winds rich in $^{26}Al$ whereas $^{60}Fe$ is ejected at a later time following the supernova explosion (Tang and Dauphas, 2012). Wolf-Rayet winds would have carved $^{26}Al$-rich bubbles in molecular cloud material, which could have subsequently been incorporated in the molecular cloud core that formed the solar system. Semi-analytic approaches have been used recently to assess the feasibility of such a scenario (Gounelle and Meynet, 2012; Young, 2014) but it remains to be seen whether high $^{26}Al$/low $^{60}Fe$ regions can exist because SN explosion is the main factor that drives the mixing between W-R product and surrounding cloud material, so that the $^{26}Al$-rich W-R material may be contaminated with $^{60}Fe$-rich supernova ejecta. Existing semi-analytic models aimed at explaining the high $^{26}Al/^{60}Fe$ ratio in meteorites have not addressed this critical aspect of the process, which will require high-resolution modeling of the interactions of stellar winds and ejecta with the surrounding medium.

## 5. Conclusion

We report Ni isotope measurements by MC-ICPMS of mineral separates from the Sahara 99555 quenched angrite (formed ~5 Myr after solar system formation) and bulk chondrules from the Semarkona LL3.00 ordinary chondrite (formed ~2 Myr after solar system formation). These two objects are important anchors in early solar system chronology. In both cases, resolvable excess $^{60}$Ni from $^{60}$Fe-decay is found. The Semarkona chondrule isochron defines an initial $^{60}Fe/^{56}Fe$ ratio of $(5.29\pm3.27)\times10^{-9}$ at the time of chondrule formation (Fig 1A). The Sahara 99555 mineral separate isochron defines and initial $^{60}Fe/^{56}Fe$ ratio of $(1.96\pm0.77)\times10^{-9}$ (Fig. 1B). These two $^{60}Fe/^{56}Fe$ add to an already long list of meteoritic materials for which the $^{60}Fe/^{56}Fe$ abundances are constrained namely bulk HEDs (Vesta), SNCs (Mars), bulk angrites, D'Orbigny angrite, Gujba (CB) chondrules, and NWA 5717 chondrules. The initial $^{60}Fe/^{56}Fe$ ratio in Sahara 99555 may have been disturbed by terrestrial alteration but all samples measured by MC-ICPMS give a consistent uniform initial $^{60}Fe/^{56}Fe$ of $(1.01 \pm 0.27)\times10^{-8}$ (Fig. 2, Table 2). This is almost two orders of magnitude lower than the estimated value from chondrule pyroxene measurements by SIMS, which give an initial $^{60}Fe/^{56}Fe$ ratio of $\sim7\times10^{-7}$. We have no explanation for the discrepancy but note that in many cases, the SIMS $^{60}Fe/^{56}Fe$ isochrons are defined by points that largely overlap with zero (Fig. 3) and were only measured on one sample type (pyroxene in chondrules from unequilibrated ordinary chondrites). In contrast, the MC-ICPMS results were measured on a variety of planetary materials formed at different times and the significances of the isochrons are in most cases very high. Until unambiguous internal isochrons are measured *in situ*, the weight of evidence favors a low $^{60}Fe/^{56}Fe$ ratio at solar system birth. Such low ratio contrasts with the high $^{26}Al/^{27}Al$ ratio, which can be explained if $^{60}$Fe was derived from the long-term chemical evolution of the Galaxy while $^{26}Al$ was derived from the wind of a Wolf-Rayet star. To test this idea, detailed modeling of the interaction of SN ejecta with W-R bubbles and surrounding medium is needed.

**Acknowledgements**

We thank V. Dwarkadas and B.S. Meyer for discussions. G.J. MacPherson (Smithsonian Institution) generously provided the Semarkona chondrules analyzed in this study. This work was supported by grants NNX12AH60G and NNX14AK09G from NASA to ND.

**Figure captions:**

**Fig.1.** $^{60}$Fe-$^{60}$Ni isochron diagrams of Semarkona chondrules and silicate minerals in Sahara 99555. Ni isotopic ratios are reported using the ε-notation; $\varepsilon^{60}Ni=[(^{60}Ni/^{58}Ni)_{sample}/(^{60}Ni/^{58}Ni)_{standard}-1]\times10^4$, where $^{60}$Ni/$^{58}$Ni ratios have been corrected for natural and laboratory-introduced mass fractionation by internal normalization to a constant $^{61}$Ni/$^{58}$Ni ratio. The error bars represent 95 % confidence intervals. In $\varepsilon^{60}$Ni *vs*. $^{56}$Fe/$^{58}$Ni isochron diagrams, the intercept gives the initial Ni isotopic composition $\varepsilon^{60}Ni_0$ while the slope is proportional to the initial $^{60}$Fe/$^{56}$Fe ratio; slope=25,961×$(^{60}Fe/^{56}Fe)_0$. Live $^{60}$Fe was detected in (A) Semarkona bulk chondrules [$^{60}Fe/^{56}Fe_i=(5.29\pm3.27)\times10^{-9}$], and (B) mineral separates of Sahara 99555 [WR, whole rock, the other labels represent fractions with different grain sizes or densities; $^{60}Fe/^{56}Fe_i=(1.96\pm0.77)\times10^{-9}$]. The green dashed line in panel A shows the expected isochron is the initial $^{60}$Fe/$^{56}$Fe ratio was $4\times10^{-7}$, as suggested by SIMS (Mishra and Goswami, 2014; Mishra and Chaussidon, 2014).

**Fig. 2.** Initial $^{60}$Fe/$^{56}$Fe ratios as a function of time after CAI formation for various meteoritic objects. The values measured by MC-ICPMS are in blue (Tang and Dauphas, 2012; this study; also see Quitté et al. 2010 for angrites). Those measured by SIMS in Semarkona chondrules are in yellow (Telus et al., 2013; Mishra and Goswami, 2014; Mishra and Chaussidon, 2014). Note the 2-orders of magnitude discrepancy between *in situ* SIMS data ($^{60}Fe/^{56}Fe=70\times10^{-8}$ at CAI formation) and those measured by MC-ICPMS ($^{60}Fe/^{56}Fe=1.01\times10^{-8}$ at CAI formation). The isochrons used to infer initial $^{60}$Fe/$^{56}$Fe ratio by SIMS are not very well defined (Fig. 3).

**Fig. 3.** Compilation of □$^{60}$Ni-$^{56}$Fe/$^{58}$Ni values measured by SIMS (Mishra and Goswami, 2014; Mishra and Chaussidon, 2014) for Semarkona chondrules in which evidence for live $^{60}$Fe was reported as significant. The reported initial $^{60}$Fe/$^{56}$Fe values for each chondrule are given in the legend (each set of colored symbols corresponds to different data points measured on the same chondrule). Panel A shows the data on a linear x-axis scale while panel B shows the data on logarithmic scale. The two black lines are internal isochrons for initial $^{60}$Fe/$^{56}$Fe ratios of $4\times10^{-7}$ (the initial value at the time of Semarkona chondrule formation reported by Mishra and Goswami, 2014; Mishra and Chaussidon, 2014) and $5\times10^{-9}$ (inferred value by MC-ICPMS, Tang and Dauphas, 2012; this study). As shown, most of the data points measured by SIMS that are taken as evidence for a high initial $^{60}$Fe/$^{56}$Fe initial ratio have uncertainties that overlap with terrestrial isotopic composition.

**Table 1.** Nickel isotopic compositions and Fe/Ni ratios of Semarkona chondrules and whole rock Sahara 99555 are from Tang and Dauphas (2012).

| Sample Name | Type | Sample Mass (mg) | Fe/Ni | $^{56}Fe/^{58}Ni$ (at.) | Norm. $^{61}Ni/^{58}Ni$ | | | Norm. $^{62}Ni/^{58}Ni$ | | | *n* Replicates |
|---|---|---|---|---|---|---|---|---|---|---|---|
| | | | | | $\varepsilon^{60}Ni$ | $\varepsilon^{62}Ni$ | $\varepsilon^{64}Ni$ | $\varepsilon^{60}Ni$ | $\varepsilon^{61}Ni$ | $\varepsilon^{64}Ni$ | |
| **Terrestrial standards** | | | | | | | | | | | |
| SRM986 | | | | | -0.036 ± 0.042 | 0.044 ± 0.170 | 0.443 ± 0.488 | -0.058 ± 0.085 | -0.033 ± 0.129 | 0.370 ± 0.453 | 12 |
| BHVO-02 | | 195 | 727 | 1010 ± 30 | -0.033 ± 0.118 | -0.040 ± 0.086 | | -0.013 ± 0.098 | 0.030 ± 0.065 | | 10 |
| BHVO-02 (2) | | 154 | | | -0.001 ± 0.074 | 0.072 ± 0.147 | | -0.037 ± 0.054 | -0.054 ± 0.111 | | 12 |
| DNC-1 | | 105 | 258 | 359 ± 23 | 0.013 ± 0.118 | 0.141 ± 0.286 | | -0.059 ± 0.091 | -0.107 ± 0.202 | | 7 |
| **Semarkona chondrules** | | | | | | | | | | | |
| SC-10-2* | Type I | 0.3 | 38 | 50 ± 2.9 | -0.03 ± 0.12 | -0.11 ± 0.17 | | 0.03 ± 0.12 | 0.09 ± 0.13 | | 10 |
| SC-30-6* | Type I | 0.2 | 13 | 17 ± 1.0 | -0.12 ± 0.25 | -0.38 ± 0.42 | | 0.08 ± 0.14 | 0.28 ± 0.32 | | 10 |
| SC-13-1 | Type II | 7.4 | 69.7 | 96.9 ± 7.9 | -0.004 ± 0.045 | 0.045 ± 0.133 | | -0.027 ± 0.065 | -0.034 ± 0.101 | | 12 |
| SC-13-2 | Type I | 3.6 | 13.2 | 18.3 ± 2.0 | -0.045 ± 0.046 | -0.011 ± 0.136 | 0.256 ± 0.643 | -0.040 ± 0.075 | 0.008 ± 0.102 | 0.272 ± 0.525 | 12 |
| SC-13-3 | Type I | 11.1 | 23.1 | 32.1 ± 2.8 | -0.026 ± 0.057 | 0.050 ± 0.096 | 0.223 ± 0.295 | -0.051 ± 0.057 | -0.038 ± 0.073 | 0.149 ± 0.326 | 12 |
| SC-13-4 | Type I | 13.5 | 20.7 | 28.8 ± 2.2 | -0.026 ± 0.045 | -0.047 ± 0.141 | 0.026 ± 0.406 | -0.003 ± 0.057 | 0.035 ± 0.106 | 0.095 ± 0.269 | 12 |
| SC-13-5 | Type I | 7.5 | 31.2 | 43.4 ± 5.2 | -0.016 ± 0.064 | 0.009 ± 0.120 | | -0.045 ± 0.046 | -0.011 ± 0.136 | | 12 |
| SC-13-6 | Type II | 1.7 | 435.0 | 605 ± 56.6 | 0.051 ± 0.043 | 0.009 ± 0.056 | | 0.047 ± 0.050 | -0.006 ± 0.042 | | 12 |
| **Sahara 99555** | | | | | | | | | | | |
| WR* | | 368 | 3376 | 4680 ± 311 | 0.31 ± 0.16 | 0.07 ± 0.22 | | 0.28 ± 0.16 | -0.05 ± 0.17 | | 6 |
| <100μm | | 121 | 3078 | 4267 ± 322 | 0.306 ± 0.090 | -0.075 ± 0.184 | | 0.344 ± 0.064 | 0.057 ± 0.140 | | 7 |
| 100-166μm | | 98 | 5613 | 7780 ± 1354 | 0.443 ± 0.134 | 0.085 ± 0.302 | | 0.400 ± 0.065 | -0.064 ± 0.228 | | 7 |
| 166-200μm | | 101 | 4512 | 6255 ± 934 | 0.310 ± 0.076 | -0.156 ± 0.177 | | 0.389 ± 0.072 | 0.118 ± 0.134 | | 10 |
| >200μm | | 131 | 5155 | 7145 ± 762 | 0.408 ± 0.048 | 0.112 ± 0.129 | | 0.351 ± 0.032 | -0.084 ± 0.097 | | 10 |
| <3.10 g/cm$^3$ | | 53 | 2538 | 3517 ± 346 | 0.214 ± 0.088 | -0.045 ± 0.173 | | 0.237 ± 0.146 | 0.034 ± 0.131 | | 7 |
| >3.10 g/cm$^3$ | | 150 | 6872 | 9524 ± 903 | 0.540 ± 0.106 | 0.087 ± 0.175 | | 0.496 ± 0.076 | -0.066 ± 0.133 | | 7 |

Note: $\varepsilon^{i}Ni = ([^{i}Ni/^{58}Ni]_{sample}/[^{i}Ni/^{58}Ni]_{SRM986}-1)\times 10^4$. The uncertainties are 95% confidence intervals.

* Nickel isotopic composition and Fe/Ni ratio in Semarkona chondrules and whole rock Sahara 99555 were from Tang and Dauphas (2012).

Table 2. Relative ages and $^{60}Fe/^{56}Fe$ ratios in different meteoritic samples and back-calculated $^{60}Fe/^{56}Fe$ initial ratio in the solar protoplanetary disk.

| Sample | Age (Myr) | Method | Reference | $^{60}Fe/^{56}Fe_i$ | Reference | $^{60}Fe/^{56}Fe_{CAI}$ |
|---|---|---|---|---|---|---|
| Bulk HED meteorites | 3.7±0.5 | $^{53}Mn$-$^{53}Cr$ anchored to D'Orbigny | Trinquier et al. (2008) | $(3.45±0.32)×10^{-9}$ | Tang and Dauphas (2012) | $(9.26±1.58)×10^{-9}$ |
| Bulk angrites | 4.9±0.2 | $^{53}Mn$-$^{53}Cr$ anchored to D'Orbigny | Shukolyukov & Lugmair (2007) | $(2.20±1.16)×10^{-9}$ | Tang and Dauphas (2012) | $(8.00±4.25)×10^{-9}$ |
| D'Orbigny minerals | 5.1±0.1 | $^{26}Al$-$^{26}Mg$ | Spivak-Birndorf et al. (2009); Schiller et al. (2010) | $(3.26±0.47)×10^{-9}$ | Quitté et al. (2010), Spivak-Birndorf et al. (2011), Tang and Dauphas (2012) | $(1.27±0.19)×10^{-8}$ |
| Sahara 99555 minerals | 5.0±0.2 | $^{26}Al$-$^{26}Mg$ | Spivak-Birndorf et al. (2009); Schiller et al. (2010) | $(1.85±0.42)×10^{-9}$ | This study, Quitté et al. (2010) | $(6.96±1.60)×10^{-9}$ |
| Semarkona chondrules | 2.0±0.8 | $^{26}Al$-$^{26}Mg$ | Kita & Ushikubo (2012) and references therein | $(5.29±3.27)×10^{-9}$ | This study | $(8.93±5.85)×10^{-9}$ |
| Gujba chondrules | 5.2±0.9 | $^{53}Mn$-$^{53}Cr$ anchored to D'Orbigny | Yamashita et al. (2010) | $<3.49×10^{-9}$ | Tang and Dauphas (2012) | |
| NWA5717 chondrules and fragments | 2±1 | Chondrule ages in UOC | Kita et al., (2005) | $<2.14×10^{-8}$ | Tang and Dauphas (2012) | |
| $^{60}Fe/^{56}Fe_{CAI}$ weighted average* | | | | | | $(1.01±0.27)×10^{-8}$ |

* Calculated using initial $^{60}Fe/^{56}Fe$ ratios at CAI formation excluding bulk angrites, which show disturbed $^{60}Fe$-$^{60}Ni$ systematics.

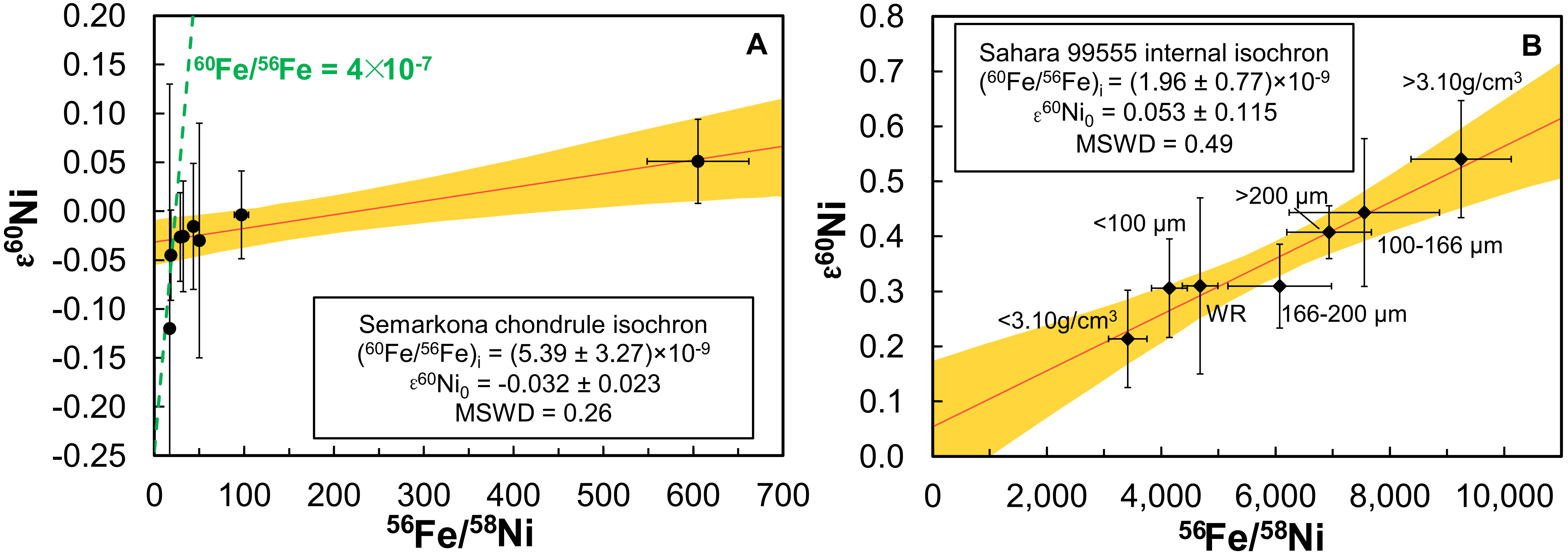

A
$^{60}$Fe/$^{56}$Fe = 4×10$^{-7}$
Semarkona chondrule isochron
($^{60}$Fe/$^{56}$Fe)$_i$ = (5.39 ± 3.27)×10$^{-9}$
ε$^{60}$Ni$_0$ = -0.032 ± 0.023
MSWD = 0.26
ε$^{60}$Ni
$^{56}$Fe/$^{58}$Ni
B
Sahara 99555 internal isochron
($^{60}$Fe/$^{56}$Fe)$_i$ = (1.96 ± 0.77)×10$^{-9}$
ε$^{60}$Ni$_0$ = 0.053 ± 0.115
MSWD = 0.49
>3.10g/cm$^3$
>200 μm
<100 μm
100-166 μm
WR
166-200 μm
<3.10g/cm$^3$
ε$^{60}$Ni
$^{56}$Fe/$^{58}$Ni

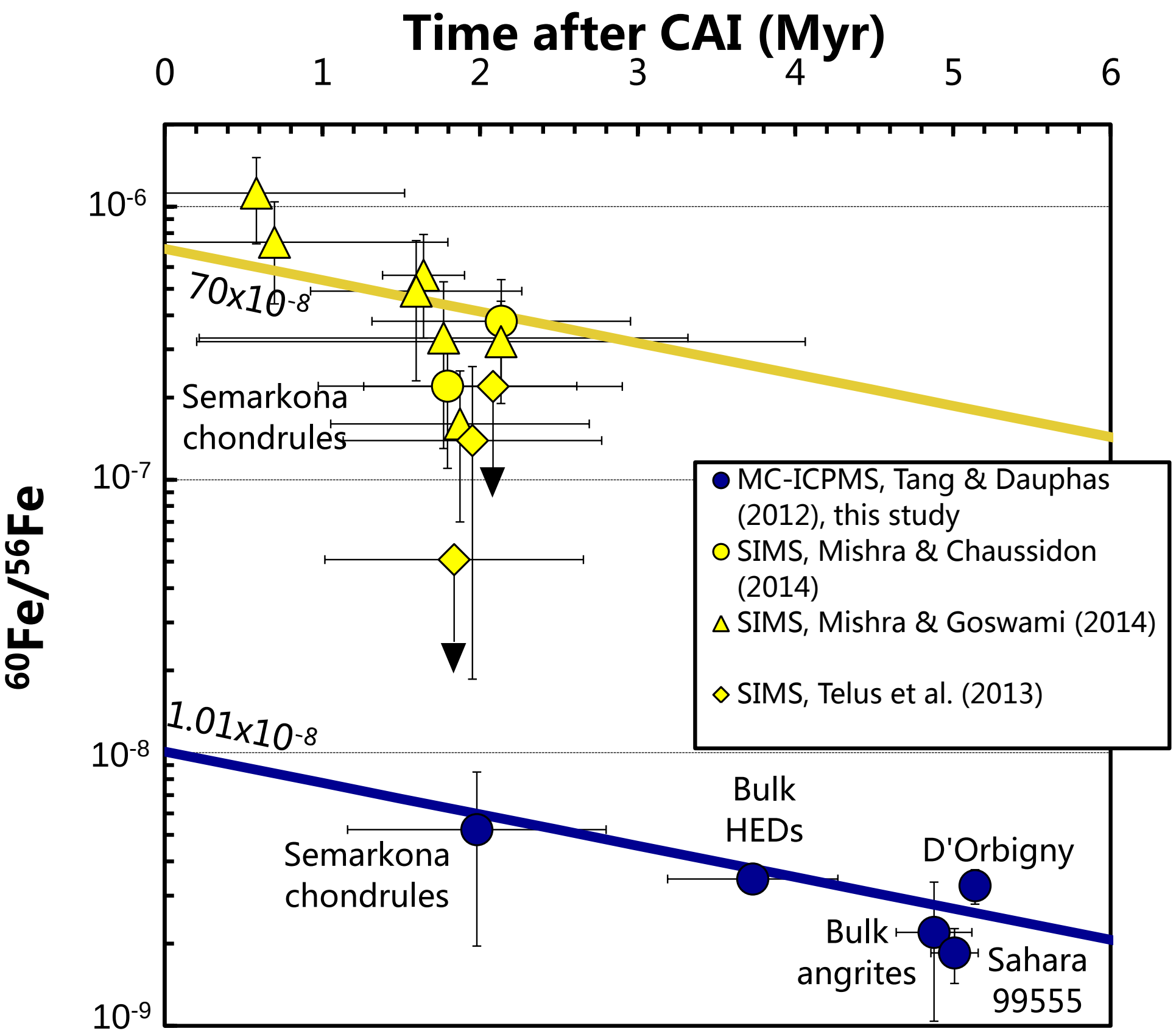

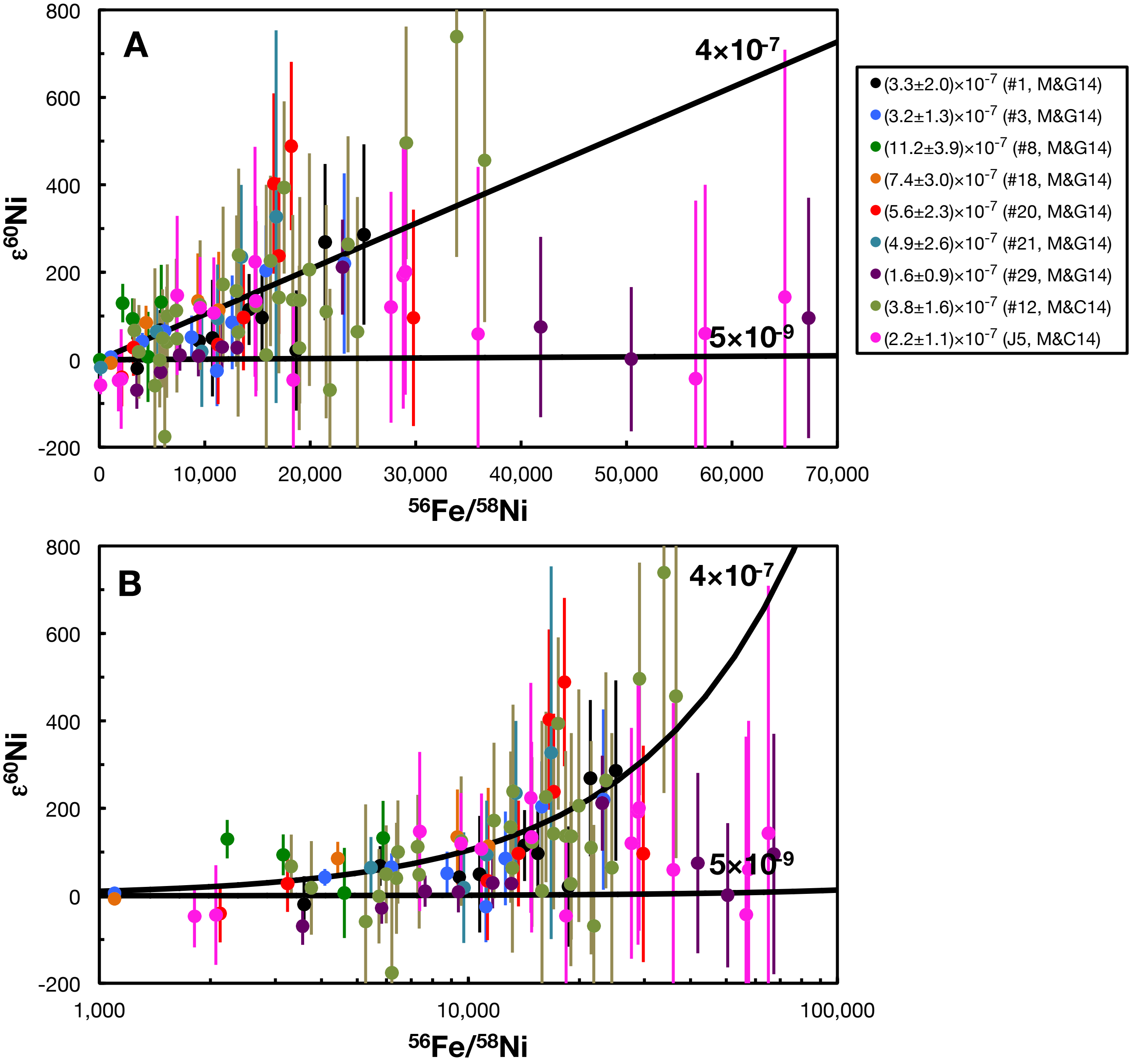